\documentclass[a4paper,11pt]{article}
\usepackage[dvips]{graphicx}
\newcommand{\lbl}[1]{\label{eq:#1}}
\newcommand{\rf}[1]{(\ref{eq:#1})}
\newcommand{\be}{\begin{equation}}
\newcommand{\en}{\end{equation}}
\newcommand{\bea}{\begin{eqnarray}}
\newcommand{\ena}{\end{eqnarray}}
\newcommand{\np}[1]{Nucl.\ Phys.\ {\bf #1}}

\newcommand{\pr}[1]{Phys.\ Rev.\ {\bf #1}}
\newcommand{\prl}[1]{Phys.\ Rev.\ Lett.\ {\bf #1}}
\newcommand{\pl}[1]{Phys.\ Lett.\ {\bf #1}}
\newcommand{\ap}[1]{Ann. Phys. (NY)\ {\bf #1}}

\newcommand{\zp}[1]{Zeit. Phys.{\bf #1}}

\def\overleftrightarrow#1{\vbox{\ialign{##\crcr
$\leftrightarrow$\crcr\noalign{\kern-1pt\nointerlineskip}
$\hfil\displaystyle{#1}\hfil$\crcr}}}
\newcommand{\lapprox}{%
\mathrel{%
\setbox0=\hbox{$<$}\raise0.6ex\copy0\kern-\wd0\lower0.65ex\hbox{$\sim$}}}

\newcommand{\gapprox}{%
\mathrel{%
\setbox0=\hbox{$>$}\raise0.6ex\copy0\kern-\wd0\lower0.65ex\hbox{$\sim$}}}

\setbox0=\hbox{$\scriptstyle\circ$}
\newcommand{\Mchir}{
\mathrel{
\setbox1=\hbox{$M$}
\copy1\kern-0.5\wd1\raise1.1\ht1\copy0 }}

\newcommand{\Gchir}{
\mathrel{
\setbox1=\hbox{$\Gamma$}
\copy1\kern-0.7\wd1\raise1.1\ht1\copy0}}

\newcommand{\Pichir}{
\mathrel{
\setbox1=\hbox{$\Pi$}
\copy1\kern-0.7\wd1\raise1.1\ht1\copy0\kern0.125\wd1}}

\newcommand{\rochir}{
\mathrel{
\setbox1=\hbox{$\rho$}
\copy1\kern-0.6\wd1\raise1.1\ht1\copy0}}

\def\pip{{\pi^+}}

\def\piz{{\pi^0}}
\def\mpi{{M_\pi}}
\def\mpid{{M^2_\pi}}
\def\mpidp{{M^2_\pip}}
\def\mpidz{{M^2_\piz}}
\def\mpiq{{M^4_\pi}}
\def\fpi{{F_\pi}}

\begin{document}
\vskip -1 truecm
\rightline{IPNO/TH 98-09}

\begin{center}
{REANALYSIS OF THE DAS ET AL. SUM RULE AND }\\
{APPLICATION TO CHIRAL $O(p^4)$ PARAMETERS}

\bigskip

Bachir Moussallam

{\sl I.P.N., Division de Physique Th\'eorique}\\
{\sl Universit\'e Paris-Sud, F-91406 Orsay C\'edex}
\end{center}

\centerline{\large\bf Abstract}

A sum rule due to Das et al. is reanalyzed using a euclidian space 
approach and a Pad\'e resummation procedure.
It is shown that the result is essentially determined by the matrix elements
of dimension six and dimension eight operators which have recently been
measured by the ALEPH collaboration. 
The result is further improved by using the vector spectral function which
must be extrapolated to the chiral limit.
This extrapolation is shown to be reliably performed under the
constraint of a set of sum rules.
The sum rule is employed not
as an approximation to $M_{\pi^+}-M_{\pi^0}$ but as an exact result for a
chiral low-energy parameter. A sufficiently precise evaluation 
provides also an estimate for a combination
of subleading electromagnetic low-energy parameters. 

\bigskip

\noindent{\large\bf 1. Introduction}

Chiral perturbation theory (e.g. \cite{chpt} for a comprehensive review) 
is now claiming to reach such a high degree of
accuracy in some situations that it is becoming necessary to deal
quantitatively with radiative corrections in low energy processes. An
important example is the pion-pion scattering amplitude for which the
two-loop contribution has recently been evaluated\cite{2pi2la}\cite{2pi2lb}.
The relevance of this reaction for probing experimentally
a basic issue in the spontaneous
breaking of chiral symmetery in QCD is discussed in some detail in 
ref.\cite{2pi2la}.
Calculations of radiative corrections have started to be performed both for
the pionium atomic bound state
(e.g. \cite{hagop} and references therein), in view of an experiment planning
to form pionium atoms at CERN\cite{nemenov}, 
and for the scattering
amplitude\cite{meissner}\cite{knecht}. The framework for performing 
such calculations is a
natural extension of the conventional chiral expansion to include the photon
as a dynamical quantum field\cite{urech}. This extension brings in a set of
new, a priori unknown, low-energy constants. At chiral order two, a single
constant appears (which will be denoted by $C$ below) while at the next
chiral order, one has to deal with fourteen new constants called $k_i$ in the
case of the $SU(2)\times SU(2)$ chiral group \cite{meissner}\cite{knecht}. 

The purpose of this
paper is to reanalyze the classic sum rule of Das et al.\cite{dgmly}.
Since experimental data has started to become available from $\tau$ decays
into hadrons, the sum rule was discussed several times in the literature
\cite{peccei}\cite{DG}\cite{aleph2} using as input experimentally measured
vector and axial-vector spectral functions. One must be cautious, however,
that the sum rule can at best provide an approximation to the $\pi^+-\pi^0$ 
mass difference if the spectral functions are not extrapolated to the chiral
limit.
Indeed, the derivation is made in the limit, $m_u=m_d=0$ and, 
stricly speaking, the integral diverges if one uses physical spectral
functions over an infinite range. 
In modern context, the sum rule must be interpreted  as an 
{\it exact} result for the low-energy
constant $C$. This constant appears in the leading term of the chiral
expansion of the $\pi^+-\pi^0$ mass difference
\begin{equation}\lbl{eq1}
M^2_{\pi^+}-M^2_{\pi^0}={2e^2C\over F^2} +O(e^2 M^2_{\pi^0})+O(
(m_u-m_d)^2)+O(e^4)\ .
\end{equation}
The corrective terms $O(e^2 M^2_{\pi^0})$ and $O( (m_u-m_d)^2)$ involve
a number of low energy constants $k_i$
and one $O(p^4)$ constant ($l_7$) respectively\cite{knecht}\cite{gl84}.
The order of magnitude of low-energy constants such as $k_i$ is known
from rather general considerations on effective theories\cite{georgi}
to be $k_i\simeq F^2_\pi/\Lambda^2$, where $\Lambda$ is the typical mass of
the massive states, not included in the effective theory, i.e.
$\Lambda\simeq M_\rho$ 
(or $M_K$, $M_\eta$ in the $SU(2)\times SU(2)$ expansion).
This enables one to estimate that the corrective terms
in eq.\rf{eq1} could be as large as $20-30\%$. 
Our claim is that by a clever use of $\tau$-decay data
recently released by the ALEPH collaboration\cite{aleph1}\cite{aleph2}
it is actually possible to
perform the sum rule evaluation of $C$ in such a precise way as to actually
provide an estimate for the combination of low-energy constants involved in
the corrective terms in eq.\rf{eq1}.

In practice, we advocate an approach in which one first constructs
the QCD correlation function $<VV-AA>$ in the chiral limit
in ${\it euclidian}$
space, an idea which was proposed in ref.\cite{bbg}. 
A key ingredient for this construction is the experimental measurement
by the ALEPH collaboration\cite{aleph2} of the vacuum matrix
elements of the dimension six and dimension eight 
combination of operators which control
the first two terms in the asymptotic expansion of the chiral correlator.
In euclidian space, far from the resonance region, this asymptotic expansion
is expected to be accurate down to rather low momenta values, say
$p\simeq 2$ GeV. The task is then to interpolate a smooth function
of $ p$, the value of which is known at zero (in terms of $F_\pi$
in the chiral limit), in a finite momentum
range. In this approach, the momentum integral
in the $[0,\infty]$ range can then be performed exactly. It will be argued
that the only knowledge of the two operator matrix elements (together with
$F_\pi$) constrains the value of $C$ to a level close to 10\%. The estimate
will then be refined by using more detailed experimental information
on the vector and the axial-vector spectral functions.

\noindent{\large\bf 2. Description of the method}

The starting point is the sum rule derived by Das et  al.\cite{dgmly}
(a quick derivation can be found in ref.\cite{moi})
written as an integral in four dimensional euclidian space. Performing the
angular integration, one expresses the constant $C$ as a one dimensional
integral 
\be\lbl{zsr}
C={3\over4}{1\over16\pi^2 }\int_0^\infty ds\, s \left[ \Pichir_A(-s)
-\Pichir_V(-s)\right]
\en
where $\Pichir_A$ and $\Pichir_V$ are defined as  the limit
when $m_u=m_d=0, e^2=0$  of the
form-factors $\Pi_A$ and $\Pi_V$ associated with the
axial-vector and the vector two-point correlation function. $\Pi_V$, 
for instance, is defined as
\be
i\int d^4x\,e^{ipx}<0\vert T V_\mu(x) V^\dagger_\nu(0)\vert 0>=
(p_\mu p_\nu-p^2 g_{\mu\nu})\Pi_V(p^2)+p^2 g_{\mu\nu}\Pi^0_V(p^2), 
\en
$V_\mu$ being the charged vector current
$V_\mu(x)=\bar u(x) \gamma_\mu d(x)$. An exactly analogous definition holds
for $\Pi_A$.

Formula \rf{zsr} is exact
provided chiral symmetry is spontaneously broken in QCD with two
massless quarks. It is  of
interest to further consider the $SU(3)\times SU(3)$
chiral limit obtained by sending
$m_s$ to zero as well. However, as will be seen in the sequel, the
uncertainties
involved in this extrapolation are too large and do not permit a useful
evaluation of $C_0=\lim_{m_s=0}C$.
Convergence of the integral in \rf{eq1} follows
from applying the operator-product expansion\cite{wilson}\cite{wbook}.
The operators must belong to the $(3,3)$ representation of the $SU(2)\times
SU(2)$ group. In the limit $m_u=m_d=e^2=0$ the only such operators that one
can construct are of dimension six or more. The spontaneous breaking of
chiral symmetry then implies that the vacuum expectation values must be
non-vanishing. 
The
following asymptotic expansion thererefore holds,
\be\lbl{expasy}
\lim_{p^2\to\infty} \Pichir_A(-p^2)-\Pichir_V(-p^2)={\lambda_6\over p^6}+
{\lambda_8\over p^8}+\ldots 
\en
In QCD, $\lambda_6$ and $\lambda_8$ are not exactly constants except at
leading order in $\alpha_s$. At higher orders, corrections carry
logarithmic-type $p^2$ dependences. 
In the leading-logarithmic approximation, this $p^2$ variation 
is found to be rather slow,
such that the approximation of taking constant 
values for $\lambda_6$ and $\lambda_8$  will be accurate 
in a reasonably large energy region. We will return to this point in sec.4. 
The parameters $\lambda_6$ and 
$\lambda_8$ have been 
determined experimentally using $\tau$ decay data\cite{aleph2}. 
The
method consists in using the analyticity properties of the two-point
functions together with Cauchy theorem, which leads to equations like
\be\lbl{cauchy}
\int_0^{M^2_\tau} ds P^{kl}(s)\left[\rho_A(s)-\rho_V(s)\right]=
{1\over2i\pi}\oint_{\vert z\vert=M^2_\tau} dz\,
P^{kl}(z)\left[\Pi_A(z)-\Pi_V(z)\right]
\en
where $P^{kl}(s)$ can be any polynomial. A convenient set is\cite{pl}
\be
P^{kl}(s)=\left(1-{s\over M^2_\tau}\right)^{k+2}
\left({s\over M^2_\tau}\right)^l
\left(1+{2s\over M^2_\tau}\right),\quad k,l>0\ .
\en
For $k=l=0$ the left hand side of eq.\rf{cauchy} reduces to a difference of
total $\tau$ decay rates. These polynomials have the further merit to
suppress the contributions which are close to the cut in the integral over
the circle so that one can use asymptotic QCD
expansions with some confidence in
the righthand side of eq.\rf{cauchy}. Using this method, the
ALEPH collaboration has determined the value of the following 
dimensionless integrals
involving $\lambda_6$ and $\lambda_8$,
\be\lbl{delta2n}
\delta^{(2n)}=-{4\pi i\over M^2_\tau}
\oint_{\vert z\vert=M^2_\tau} dz P^{00}(-z){\lambda_{2n}(z)\over z^n},\quad
n=3,\ 4 
\en
to be\cite{aleph2},
$\delta^{(6)}=-0.058\pm0.006$ and $\delta^{(8)}=0.0170\pm0.0014$. Ignoring
the $p^2$ dependence of $\lambda_6$ and $\lambda_8$ (the validity of this
approximation in the present situation  can be 
checked explicitly for $\lambda_6$ and will be
found in sec.4 to be excellent) one deduces,
\be\lbl{l6l8=}
\lambda_6=(7.58\pm0.80) 10^{-3} \ {\rm GeV}^6\qquad
\lambda_8=(-1.07\pm0.12) 10^{-2} \ {\rm GeV}^8\ .
\en
The result for $\lambda_6$ is in reasonable agreement with that obtained
earlier \cite{lamb6}. An analysis of the ALEPH data, making use of negative
moments performed very
recently\cite{luca} leads to values compatible with \rf{l6l8=} although
slightly smaller.
It is perhaps important to stress that, even
though $\lambda_6$ and $\lambda_8$ control the expansion of a chiral limit
correlator they are effectively correctly determined from data in which
$m_u, m_d\ne0$. This is because quark mass effects
are properly taken into account in the fit as they occur in
the operator-product expansion via operators of lower dimensionality and the
contribution of dimension six linear in the quark mass ( involving
the so-called mixed
condensate) happens to vanish at leading order in $\alpha_s$\cite{svz}. In
other terms, the chiral correction to $\lambda_6$ is strongly suppressed.
A priori, there is no reason for a similar suppression to hold for
$\lambda_8$, but this parameter is of lesser practical importance in the
calculation. A key assumption which is made in the above determination of
$\lambda_6$ and $\lambda_8$ concerns, of course, the validity of truncating
the asymptotic expansion \rf{expasy} at order eight for $p=M_\tau$. We
will see below that this assumption is internally consistent but it is not
easy to estimate the error induced by this truncation. For this purpose, one
should be able determine more asymptotic parameters and check the stability
of the determination.

Let us now explain the method for evaluating the integral in eq.\rf{zsr}. We
first split the integrand in two parts
\be\lbl{pisplit}
\Pichir_A(-s)-\Pichir_V(-s)=\Pi^{exp}_{A-V}(-s)+\Pi^{rem}_{A-V}(-s)\ ,
\en
where $\Pi^{exp}_{A-V}(-s)$ is constructed from an experimentally measured
part of the vector and the axial-vector spectral functions. We will proceed
in three successive steps of approximation, including more and more
experimental information in this part, and then check the stability of the
result. In the first approximation, we include solely the pion pole part,
\be\lbl{1st}
\Pi^{exp}_{A-V}(-s)={2F^2\over s}\quad(1^{st} \ approximation)
\en
where $F$ is the pion decay constant $F_\pi\simeq92.4$ (MeV) extrapolated
to the chiral limit. The remainder part in eq.\rf{pisplit},
$\Pi^{rem}_{A-V}$, is reconstructed
from its asymptotic expansion assuming that four terms in this expansion are
known
\be\lbl{asyexp}
\lim_{s\to\infty}\Pi^{rem}_{A-V}(-s)={a_2\over s}+{a_4\over s^2}+
{a_6\over s^3}+{a_8\over s^4}+\ldots \ .
\en
For instance, in the first order approximation corresponding to \rf{1st},
one would have $a_2=-2F^2$, $a_4=0$, $a_6=\lambda_6$, $a_8=\lambda_8$. The
point is that, firstly, we expect this asymptotic expansion to become
numerically accurate at rather low values of the momenta, 
$\sqrt{s}\simeq2$ GeV. 
Secondly, the function
$\Pi^{rem}_{A-V}(-s)$ is expected to be a perfectly smooth function down to
$s=0$. In the first order approximation, it has
a logarithmic chiral singularity at $s=0$ 
with a small numerical coefficient,
\be\lbl{log}
\lim_{s\to0}\Pi^{rem}_{A-V}(-s)={1\over24\pi^2}\log s+cstt
\quad(1^{st}\ approximation)\ .
\en
In higher order approximations this singularity will be exactly included in
$\Pi^{exp}_{A-V}(-s)$ and the remainder part will be finite at $s=0$.
It is plausible that a simple
rational approximation should be able to interpolate rather precisely the
remainder function in the range $\sqrt{s}=[0,2]$ GeV. Imposing finiteness at
$s=0$ and matching to the asymptotic expansion 
selects a unique kind of Pad\'e approximant,
\be\lbl{pade}
\Pi^{rem}_{A-V}(-s)={a s +b\over s^2 +c s +d}\ .
\en
The parameters of the approximant being related to those
occuring in the asymptotic expansion eq.\rf{asyexp} by the simple relations
\be
d={a^2_6-a_4 a_8\over a^2_4-a_2 a_6},\quad
c={a_2 a_8-a_4 a_6\over a^2_4-a_2 a_6},\quad
b=a_4+a_2 c,\quad
a=a_2\ .
\en
In the second level of approximation we include into $\Pi^{exp}_{A-V}(-s)$
the most significant part of the $2\pi$ spectral function together with the
one-pion pole which was considered before
\be
\Pi^{exp}_{A-V}(-s)={2F^2\over s}-\int_0^{M^2_\tau}{\rochir_{2\pi}(x)
\over x+s}dx \quad (2^{nd}\  approximation) .
\en
In this approximation, the logarithmic singularity \rf{log} is
properly taken into account provided the spectral function is correctly
normalized at the origin: $\rochir_{2\pi}(0)=1/24\pi^2$\cite{gl84}. The
construction of the chiral limit spectral function $\rochir_{2\pi}$ knowing
the experimentally measured one $\rho_{2\pi}$ is not a completely trivial
matter and will be explained in the next
section. The remainder piece is constructed as a Pad\'e approximant as before
except that the asymptotic expansion parameters $a_i$ which enter are now
given by
\be
a_2=-2F^2+I_0,\quad a_4=-I_1,\quad a_6=\lambda_6+I_2,\quad
a_8=\lambda_8-I_3
\en
with
\be
I_n=\int_0^{M^2_\tau} dx\,x^n\rochir_{2\pi}(x)\ .
\en

One can of course think of continuing in this way and include more and more
experimental information such that the remainder function will become
numerically smaller together with the uncertainty associated with the Pad\'e
interpolation procedure. The next step, then, 
would be to include explicitly the contribution from the
three pion component of the spectral function,
\be
\Pi^{exp}_{A-V}(-s)={2F^2\over s}-
\int_0^{M^2_\tau}{\rochir_{2\pi}(x)\over x+s}dx+
\int_0^{M^2_\tau}{\rochir_{3\pi}(x)\over x+s}dx
\quad (3^{rd} \ approximation) .
\en
What prevents one from pursuing this construction further lies in the
difficulty of performing the chiral extrapolation, which increases with the
pion multiplicity. It will fortunately appear that convergence is very fast,
such that one hardly needs to go beyond the second approximation.

\noindent{\large\bf 3. Chiral limit extrapolations}

\noindent{\bf 3.1 $F_\pi$}

Extrapolation to the chiral limit of $F_\pi$ 
can be performed fairly easily using known results from 
chiral perturbation theory. The value of $F$,
corresponding to $m_u=m_d=0$, $m_s\ne0$ is related to $F_\pi$ at one loop
order by the following expression\cite{gl84}
\be\lbl{Fdev}
F=F_\pi\left(1-{13\over192\pi^2}{\mpid\over F^2_\pi}-{\mpid\over6}
\langle r^2\rangle^\pi_S 
+O(M^4_\pi)\right)\ ,
\en
which involves the scalar radius of the pion.  
Using for this quantity the updated value as given in ref.\cite{dafne}: 
$\langle r^2\rangle^\pi_S=0.60\pm0.05\ {\rm fm}^2$, one obtains
\be\lbl{F}
F=86.7\pm0.6\pm 0.5\ {\rm MeV}
\en
using $F_\pi=92.4\pm0.3$ MeV\cite{pdg96}. The second error in the value of
$F$ is a naive 
order of magnitude estimate of the size of the $O(\mpiq)$ correction in 
eq.\rf{Fdev}. 
This relatively precise
extrapolation is to be contrasted with the situation in which one would be
willing to further extrapolate to $m_s=0$. Let $F_0$ be the corresponding
limiting value of $F_\pi$, it is related to $F$ by the following relation
\cite{gl85}
\be
F_0=F\left(1-{8m_sB\over F^2} L^r_4(m_sB) +O(m_s^2)\right)\ ,
\en
where $B$ is proportional  to the quark condensate in the chiral limit, 
$B=-<\bar u u>/F^2$. 
This relation involves the low-energy constant $L_4$. 
Unfortunately, there is  no independent way of determining
$L_4$, which appears here multiplied by a large numerical factor.

\noindent{\bf 3.2 $\rho_{2\pi}$}

Let us now discuss the two-pion component of the vector spectral function.
The possibility of performing a reliable extrapolation here is tied to the
time old observation of vector meson dominance of the pion
electromagnetic form-factor $F_V$. Defining $\rho_{2\pi}$ in
terms of  $F_V$,
\be
\rho_{2\pi}(s)=\theta(s-4M^2_\pi)\,
{1\over24\pi^2}\left({s-4M^2_\pi\over s}\right)^{3\over2}
\vert F_V(s)\vert^2\ ,
\en
an excellent fit to the data can be performed 
up to the tau meson mass, with a
Breit-Wigner function for the $\rho$ resonance and only a small admixture of
higher mass resonances,
\be
F_V(s)={1\over1+\beta+\gamma}\left(B_\rho(s)+\beta B_{\rho'}(s)+
\gamma B_{\rho''}(s)\right)\ ,
\en
with
\be\lbl{bwro}
B_\rho(s)={M^2_\rho\over M^2_\rho-s-i\sqrt{s}\,\Gamma_\rho(s)},\quad
\Gamma_\rho(s)=\theta(s-4M^2_\pi)\,\Gamma_\rho\,{M^2_\rho\over s}
\left({s-4\mpid\over M^2_\rho-4\mpid}\right)^{3/2}\ .
\en
This type of parametrization guarantees that $F_V(0)=1$ and was proposed in
ref.\cite{ks}. We will be using the numerical values obtained from a
combined fit of the ALEPH $\tau\to2\pi$ decay data and the $e^+ e^-\to
\pi^+\pi^-$ data\cite{aleph1}
\bea\lbl{paramv}
&&M_\rho=773.4\pm0.9,\quad \Gamma_\rho=147.7\pm1.6,
         \quad \beta=-0.229\pm0.020\\ \nonumber
&&M_{\rho'}=1465\pm22,\quad \Gamma_{\rho'}=696\pm47,	 
         \quad \gamma=0.075\pm0.022\\ \nonumber
&&M_{\rho''}=1760\pm31,\quad \Gamma_{\rho''}=215\pm86\ .
\ena
Other variants in the functional form of the Breit-Wigner function
$B_\rho(s)$
may be used which would result in somewhat different values of the parameters
\rf{paramv}. In particular, the form due to Gounaris and Sakurai\cite{gs} has
better analytical properties and can
approximately correctly reproduce the cut of $F_V(s)$ in the
chiral limit while the simpler form \rf{bwro} produces no cut at all. 
Nevertheless, for the
problem at hand, we found numerically insignificant differences in using
either parametrization.

It is clear that extrapolation to the chiral limit will dominantly 
affect the lower
energy part of the spectral function. Furthermore, the uncertainties in the
parameters of the higher resonances $\rho'$, $\rho''$ are larger than the
effect of setting $m_u=m_d=0$. 
Therefore, in order to obtain the spectral function 
in the chiral limit it is only necessary to
evaluate the extrapolated values of the $\rho$-meson mass and width,
$\ \Mchir_\rho$ and $\Gchir_\rho$. Let us now discuss this issue.

In the case of the mass, firstly, one can perform a chiral expansion. At
leading order, linear in the quark masses, the $\rho$ and $K^*$ masses 
are expressed in terms of 
two independent parameters (besides $\Mchir_\rho$ ) $B_1$ and $B_8$,
\be
M_\rho=\Mchir_\rho +2\hat m B_8 + 2\hat m B_1\qquad
M_{K^*}=\Mchir_\rho +(m_s+\hat m )B_8 + 2\hat m B_1\ .
\en
One needs in principle to know both of these parameters in order to deduce
$\Mchir_\rho$. 
The parameter $B_8$ is easily obtain $K^*-\rho$ mass difference,
\be\lbl{B8}
2\hat m B_8={2(M_{K^*}-M_\rho)\over r-1},
\quad r={m_s\over \hat m}\simeq 26
\en
(using the standard chiral expansion 
framework for evaluating the ratio $m_s/\hat m$).
The value of $B_1$, on the other hand, cannot be simply determined, but this
parameter is suppressed in the large $N_c$ limit and thus should be smaller
than $B_8$. Neglecting $B_1$ gives $\Mchir_\rho-M_\rho\simeq -10$ MeV. The
chiral expansion of the vector meson masses has been pursued recently beyond
linear order \cite{jmw}\cite{bgt}. Including the leading correction, which
are of order $O(m_q^{3/2})$\cite{jmw} gives for $\rho$-meson mass in 
the form,
\be
M_\rho=\Mchir_\rho +2\hat m(B_8+B_1)-{g^2\mpid\over48\pi F^2_\pi}
\Mchir_K(3+{4\over3\sqrt3}) -{g^2 M^3_\pi\over12\pi F^2_\pi}\ .
\en
Here, the parameter $g$ can be determined approximately to be $g\simeq0.60$
\cite{DW}\cite{bgt} and $\Mchir_K=\lim_{m_u,m_d=0}M_K$. The parameter $B_8$
can, again, be determined from the $K^*-\rho$ mass difference and one finds
that its numerical value is essentially the same as in the linear expansion.
The corrective terms, even though suppressed in the large $N_c$ limit, turn
out to be relatively large and approximately cancel the contribution
proportional to $B_8$. Further corrections of order $O(m_q^2)$
are also generated at one-loop which were computed in ref.\cite{bgt}. This
contribution depends on a rather large number of parameters. We will not
attempt to take it into account quantitatively but simply use the
qualitative fact that it goes in the sense of reducing somewhat the large
effect of the $O(m_q^{3/2})$ contribution, such that one can estimate with
some confidence that the $\rho$ mass in the chiral limit should lie in a
range,
\be
-10\ {\rm MeV}\lapprox\, \Mchir_\rho-M_\rho\lapprox 0\ .
\en

Concerning the chiral limit of the width of the 
$\rho$-meson, we may also try to
follow a similar approach and expand to linear order in the quark
masses. Unfortunately, even at such a low order and dropping Zweig rule
violating terms, there still remains too many undetermined constants. 
The most general chiral lagrangian terms  describing vector meson
coupling to pseudo-Goldstone boson pairs  (using notations as in
ref.\cite{egpr}) linear in the quark mass matrix are
\be\lbl{lagvpp}
{\cal L}_{vpp}={iG_V\over\sqrt2}\left(
<V_{\mu\nu}u^\mu u^\nu> +
\gamma_1<\{\chi^{(+)},V_{\mu\nu}\}u^\mu u^\nu>+
\gamma_2<V_{\mu\nu}u^\mu\chi^{(+)}u^\nu>\right)\ .
\en
Note that wave-function renormalization effects of either the chiral fields
or the vector meson fields can effectively be absorbed into the parameter
$\gamma_1$.
It turns out not to be possible to determine the three constants
$G_V$ (which determines the chiral limit width) and $\gamma_1$, $\gamma_2$
independently. Qualitatively, at least, this approach suggests, from the
phase-space factor and the pion momentum dependence of the decay matrix
element, that one should expect an increase of the $\rho$-meson width in the
chiral limit of the order of 20\%. This is a rather large effect and it
must be properly taken into account.

As a way out of these difficulties, one may construct a set of 
sum rules involving the
difference of the spectral functions $\rho_V-\rochir_V$. To the extent that
the lower part of the integration region dominates, such sum rules will
efficiently constrain the chiral limit of the $\rho$-meson parameters.
One derives a first sum rule by 
considering the combination of $\Pi_V(-s)$ minus its chiral limit
counterpart 
$\Pichir_V(-s)$. Asymptotically, one has (e.g. \cite{bnp}),
\bea
&&\lim_{s\to\infty} s\left(\Pi_V(-s)-\Pichir_V(-s)\right)=\\ \nonumber
&&\lim_{s\to\infty}{-3\over8\pi^2}\left\{
(1+{8\alpha_s(s)\over\pi})(m_u(s)+m_d(s))^2+
(1+{2\alpha_s(s)\over\pi})(m_u(s)-m_d(s))^2\right\}=0\ .
\ena
Hence, using a spectral representation, the following sum rule must hold,
\be\lbl{sr1}
\int_0^\infty dx\,\left(\rho_V(x)-\rochir_V(x)\right)=0\ .
\en
A second second sum rule, with even better convergence properties, is
obtained by considering the following $s=0$ limit,
\be\lbl{diff2}
\Pi_V(0)-\lim_{s\to0}
\left(\Pichir_V(s)+{1\over24\pi^2}\log{-s\over\mu^2}\right)
\en
This expression can be evaluated in two different ways. Firstly, one can use
the chiral expansion of the vector correlation function: a very
good level of precision can be reached  thanks to the calculation 
at two-loop order by Golowich and Kambor\cite{gk}. Secondly, one
can write down a spectral representation: here it is convenient to split the
integration range into $[0,4\mpi^2]$ and $[4\mpi^2,\infty]$. In the first
range, the integral can be performed explicitly, using the one-loop
expression for the spectral function $\rochir_V$. Equating these two
evaluations, one derives the
second sum rule,
\bea\lbl{sr2}
&&\int_{4\mpi^2}^\infty dx\,{\rho_V(x)-\rochir_V(x)\over x}=
{1\over12\pi^2}\left(\log2-{4\over3}\right)\\ \nonumber
&&\quad+{\mpid\over288\pi^4 F^2}\left( \bar l_6 -\log4+{8\over3}
+{3\over2}\,(\bar l_5-\bar l_6)\Big[\log{\mu^2\over\mpi^2}
+{1\over4}\log{\mu^2\over M^2_K}-{1\over4}\Big]\right)\\ \nonumber
&&\quad-{8\mpid\over F^2}\Big[ Q(\mu^2)+2R(\mu^2) 
+{F^2\over768\pi^2M^2_K}\Big]
+O(M^4_\pi)\ .
\ena
In this expression, $\bar l_5$ and $\bar l_6$ are low-energy constants which
appear at $O(p^4)$\cite{gl84} and which are well determined, while
$R(\mu^2)$ and $Q(\mu^2)$ are $O(p^6)$ constants\cite{gk}
(the appearance of $M_K$ in the above expression is related to the fact that
these constants are appropriate for the three-flavour chiral expansion).
One expects
$R(\mu^2)$ to be suppressed compared to $Q(\mu^2)$ because of the Zweig
rule (for values of the scale $\mu$ of the order of 1 GeV) 
and the latter constant was evaluated from a sum rule\cite{gk2}
\be
Q(M^2_\rho)=(3.7\pm 2.0) 10^{-5}\ .
\en
This enables one to evaluate the {\rm entire} $O(\mpid)$ contribution on the
right-hand side of eq.\rf{sr2}.  
The set of two sum rules \rf{sr1} and \rf{sr2} can be considered as a set of
non linear equations from which one can determine $\Mchir_\rho$ and
$\Gchir_\rho$. We have analyzed this system numerically, and found that it
has a  solution, which is unique  in a physically meaningful range. 
Corresponding to the central values of the parameters cited above 
and including only the two-pion component of the vector spectral functions, 
one obtains
\be
\Mchir_\rho-M_\rho=-2.4\ {\rm MeV}\qquad 
\Gchir_\rho=180.8\ {\rm MeV}\ .
\en
The uncertainties in this result come from
two sources.
Firstly, there is an uncertainty in the integrals of
$\rho_V$ coming from experimental errors in the parameters describing
$\rho_V$. Varying the parameters in
\rf{paramv} one finds that, essentially, the error on $M_\rho$ and
$\Gamma_\rho$ are the only ones that matter and that they translate into
identical errors on $\Mchir_\rho$ and $\Gchir_\rho$. 
Secondly, we have neglected in the integrals the
contribution of components in the vector spectral
function other than $2\pi$, i.e.  
$\rho_V^{4\pi},\ \rho_V^{K\bar K},\ \rho_V^{6\pi},\ldots $ 
Evidently, one expects the first of the sum rules to be more sensitive to
these contributions which set up at higher energies. 
One can make
a rough estimate of the influence of these components 
using the quark-hadron duality 
idea, i.e. modelling the sum of all contributions by a continuum,
\be
\rho^{cont}_V(s)={1\over4\pi^2}\theta(s-M^2_{cont})
\en
normalized to the asymptotic QCD prediction and
starting at some threshold mass $M_{cont}$. A typical value
used in sum rules analysis is $M_{cont}\simeq1.5$ GeV.
For the problem at hand, we need
to know also how this continuum mass varies when going to the chiral limit.
There is of course no way to precisely evaluate that, but it seems not
unreasonable to assume $-10\lapprox M_{cont}-\Mchir_{cont}\lapprox 10$ MeV,
which leads to a variation $\Delta\Mchir_\rho=\pm 8$ MeV. The conclusion
is that the chiral mass is, in fact, not determined to a better
accuracy from the sum rules
than it was from the chiral expansion as discussed above. Imposing
that the sum rule result be the same as that found before, i.e.
$\Mchir_\rho-M_\rho=-5\pm5$ MeV, is achieved by taking the
continuum mass parameter in the range
$\Mchir_{cont}-M_{cont}=2\pm6$ MeV. Solving
the two sum rule equations simultaneously yields the chiral mass and
width as approximately linear functions of $\Mchir_{cont}-M_{cont}$ and they
are found to lie in the range
\be
\Mchir_\rho-M_\rho=-5\pm5\ {\rm MeV}
\qquad \Gchir_\rho=180.0\pm1.5\ {\rm MeV}\ .
\en
One observes that the width gets determined with a much  smaller 
error than the mass. The
spectral function $\rho_{2\pi}$ and its chiral extrapolation are shown in
Fig. 1. From this figure one observes, in particular, that the influence of
setting $m_u$, $m_d$ to zero is felt mostly  in the low-energy region, 
$\sqrt{s}\le 1$ GeV, consistently with the starting point assumption.

\begin{figure}[ht]
\begin{center}
\title{\bf Figure 1}
\includegraphics*[scale=1]{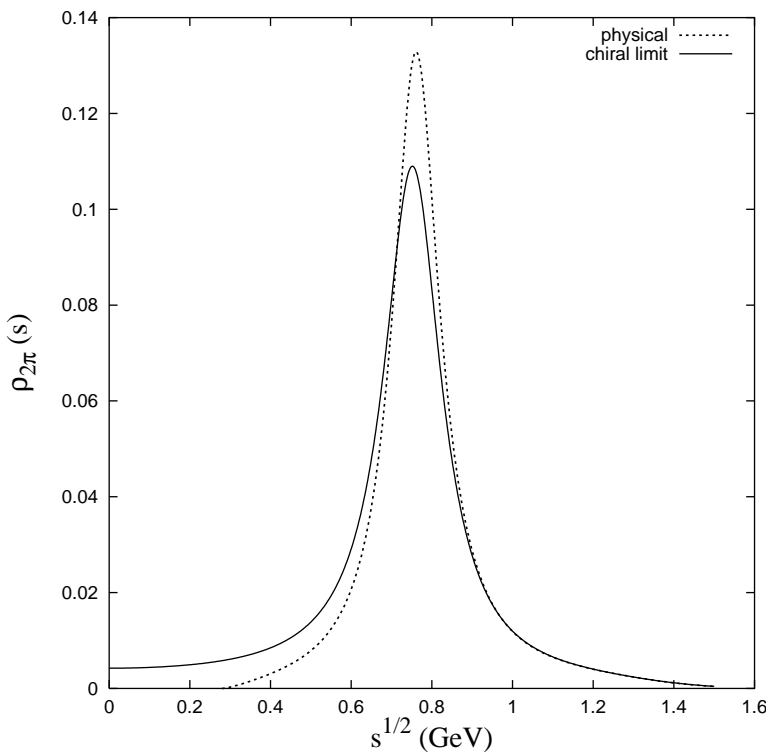}
\caption{\sl Two-pion component of the vector spectral function and its
chiral limit extrapolation obtained from solving the non-linear system of
two sum rule equations, as explained in the text.
}
\end{center}
\end{figure}

\noindent{\bf 3.3 $\rho_{3\pi}$ }

The spectral function piece
$\rho_{3\pi}$ is not known to the same accuracy as $\rho_{2\pi}$.
Furthermore, it will appear that extrapolation to $m_u=m_d=0$ is
plagued with  larger uncertainties. However,
$\rochir_{3\pi}$ is not an essential ingredient, its explicit
inclusion turns out to have very little effect and only serves
to verify the stability of the calculation. For this purpose, an
approximate knowledge of $\rochir_{3\pi}$ may be sufficient.
As before, one expects a sizable contribution from
a resonance, the $a_1(1260)$ in this case.
However, because the $a_1$ has a larger
mass than the $\rho$ and especially because it has a much larger width it is
more questionable that the background contribution will be negligible. We
will anyway follow the model of K\"uhn and Santamaria\cite{ks} which assumes
complete dominance of the $a_1$ and matches with the correct chiral
$O(p^2)$ behaviour of the axial current matrix element at low energy (
note that the
$O(p^4)$ expression has been recently worked out \cite{col}). One assumption
in this model is that the $a_1$ decays via a two step process:
$a_1\to\rho\pi\to3\pi$
or $a_1\to\rho'\pi\to3\pi$ with a small probability .
In principle, nothing prevents
the $a_1$ decay to proceed also 
via the $a_1\to\sigma\pi$ channel\footnote{We approximate, as
usual, a strongly interacting pion pair in an S-wave by a fictitious or real
but very wide $\sigma$ meson.}. A clear signature for this process
would be a difference in the $a_1$ decay rates into $2\pi^-\pi^+$ and
$2\pi^0\pi^-$.  
These two rates have now been measured separately
for the first time by the ALEPH collaboration\cite{aleph2} and found to be
equal to a very good precision 
($R^{--+}=9.1\pm0.2\%,\ R^{00-}=9.2\pm0.2\%$). This measurement
supports the decay model of ref.\cite{ks}. 
This model is embodied in the following parametrization
\be
<\pi^-(q_1)\pi^-(q_2)\pi^+(q_3)\vert\bar u\gamma^\mu\gamma^5 d\vert0>=
-i{2\sqrt2\over3F} B_{a_1}(s)\left(B_{\rho\rho'}(s_2) V_1^\mu+
B_{\rho\rho'}(s_1) V_2^\mu\right)
\en
with
\be
V_i^\mu=q_i^\mu-q_3^\mu-Q^\mu\,{Q.(q_i-q_3)\over s},\quad
Q=q_1+q_2+q_3,\quad
s_i=(Q-q_i)^2
\en
and 
\be
B_{\rho\rho'}(s_i)={B_\rho(s_i)+\beta' B_{\rho'}(s_i)\over1+\beta'},\quad
B_{a_1}(s)={M^2_{a_1}\over M^2_{a_1}-s-iM_{a_1}\Gamma_{a_1} g(s)/g(M^2_{a_1})}
\en
where $g(s)$ is a three-body phase-space integral which must be computed
numerically (see ref.\cite{ks} for more details
\footnote{An approximate analytical form of $g(s)$ is given in this
reference but one must be careful that it is only valid for the physical
value of $M_\pi$ and becomes incorrect for $M_\pi=0$.}).
We have determined the $a_1$ mass and width as well as the decay parameter
$\beta$ from a simple-minded fit of the ALEPH data\footnote{The data can be
found on the website {\sl http://alephwww.cern.ch/ALPUB/paper/paper98/1} .}
\cite{aleph2} assuming energy independent errors.
The resulting values for the $a_1$ parameters obtained in this way are,
\be
M_{a_1}=1.28\pm0.01\ {\rm GeV},\quad 
\Gamma_{a_1}=0.67\pm0.05\ {\rm GeV},\quad
\beta'=-0.27\pm0.03\ .
\en
The experimental invariant mass distribution for the mode
$\tau\to\pi^-\pi^-\pi^+\nu$ is shown in Fig. 2 together with the result of the
fit using the above parametrization. 

\begin{figure}[ht]
\begin{center}
\title{\bf Figure 2}
\includegraphics*[scale=1]{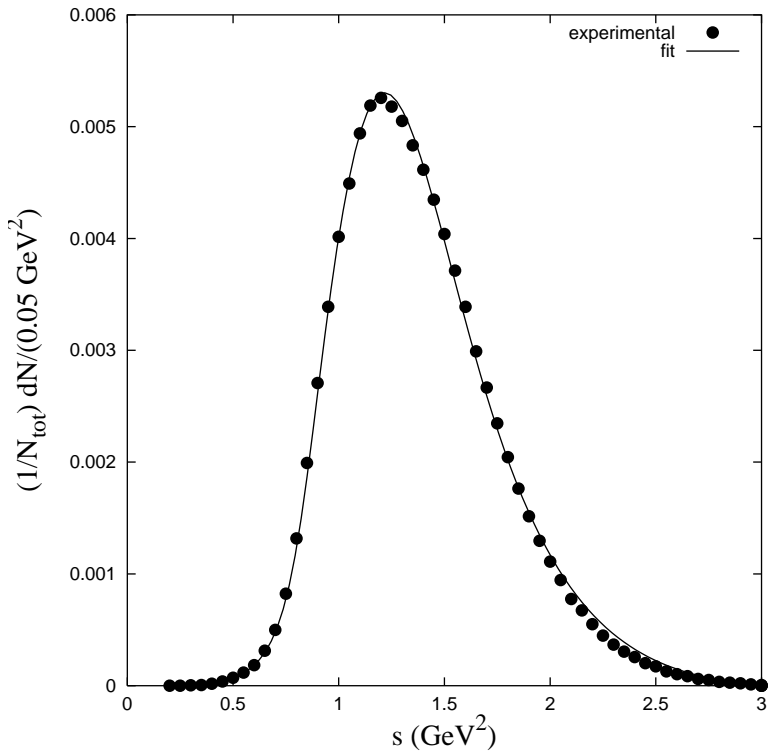}
\caption{\sl
Branching fraction for the mode  $\tau\to\pi^-\pi^-\pi^+\nu$ as a function
of the three-pion invariant mass squared. The experimental results from
ALEPH are displayed together with our fit based on the K\"uhn-Santamaria
parametrization.
}
\end{center}
\end{figure}

Now we would like to construct the chiral limit extrapolation of the $3\pi$
spectral function. As before, we disregard the modification of the parameters
associated with the $\rho'$ as it makes a relatively minor contribution
to the spectral function. Concerning the $\rho$ meson, the extrapolation of
its mass and width were discussed in the previous subsection, there
essentially remains to estimate the modification of the $a_1$ mass and width
parameters.
Concerning
the mass, one encounters the first difficulty that the quark mass matrix not
only shifts the $1^{++}$ multiplet but also mixes the states with non-zero
strangeness with those of the $1^{+-}$ multiplet.
Expanding to linear order in the quark masses, assuming ideal mixing, and
using the $s\bar s$ member of the multiplet gives,
\be
\Mchir_{a_1}\simeq M_{a_1}-{M_{f_1(1510)}-M_{a_1}\over ( r-1)}\ .
\en
This estimate must be considered as very approximate because of the
additional problem that the assignment of the $f_1(1510)$ as the $s\bar s$
member of the $a_1$ nonet\cite{pdg96} is far from certain\cite{close}. 

The value of the $a_1$ width, finally, in the chiral limit is constrained by
a sum rule exactly analogous to \rf{sr1} in the axial channel,
\be
\int_0^\infty dx\,\left(\rho_A(x)-\rochir_A(x)\right)=2F^2-2\fpi^2\ .
\en
In this equation the one-pion component is excluded from $\rho_A$ and its
contribution appears on the right-hand side. As before, the additional
assumption must be made that this equation constrains mostly the
low energy part of the spectral function and, as a consequence, can
essentially be interpreted as an equation for the $3\pi$ component of
$\rho_A$. Using $\Mchir_{a_1}-M_{a_1}=-10$ MeV, the sum rule gives
the chiral width: $\Gchir_{a_1}=0.70$ GeV.
We shall be content with a single sum rule here even though it is 
possible in principle to exploit a second sum rule in analogy with the
case of the $\rho$ meson. 
The result for the physical 
three-pion spectral function and its chiral limit is displayed in Fig.3.

\begin{figure}[ht]
\begin{center}
\title{\bf Figure 3}
\includegraphics*[scale=1]{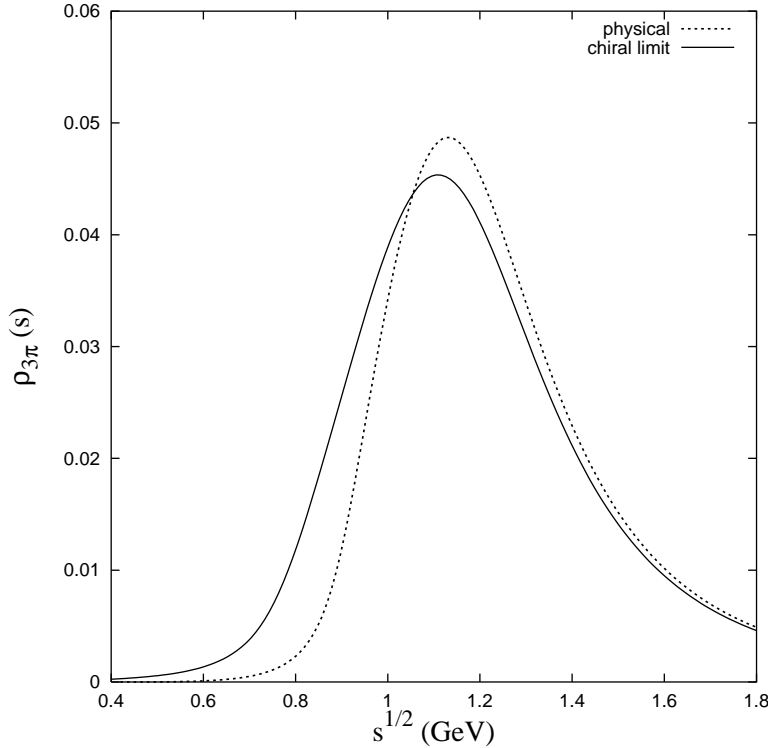}
\caption{\sl Three-pion component of the axial-vector spectral function and its
chiral limit extrapolation.
}
\end{center}
\end{figure}

\noindent{\large\bf 4. Results}

Now that we have defined an approximation scheme for
calculating $\Pi_{A-V}^{exp}$ and $\Pi_{A-V}^{rem}$ it is straightformward
to compute the sum rule integral, eq.\rf{zsr}. Before we do so, it is
instructive to have a look at the integrand, which is displayed in Fig.4 and
Fig.5 for the three levels of approximation. 

\begin{figure}[ht]
\begin{center}
\title{\bf Figure 4}
\includegraphics*[scale=1]{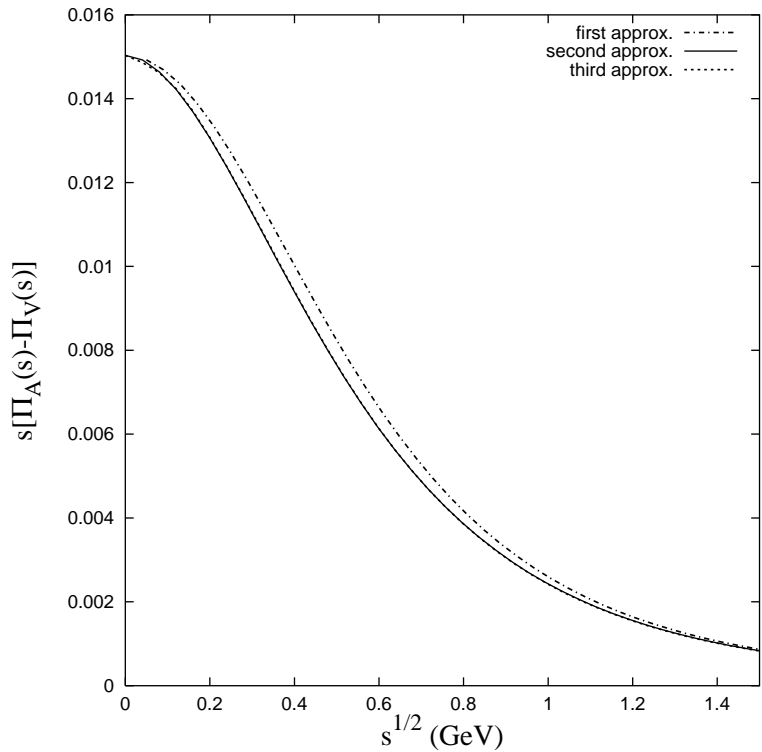}
\caption{\sl Integrand of the Das et al. sum rule in the three successive
approximations.
}
\end{center}
\end{figure}
\begin{figure}[ht]
\begin{center}
\title{\bf Figure 5}
\includegraphics*[scale=1]{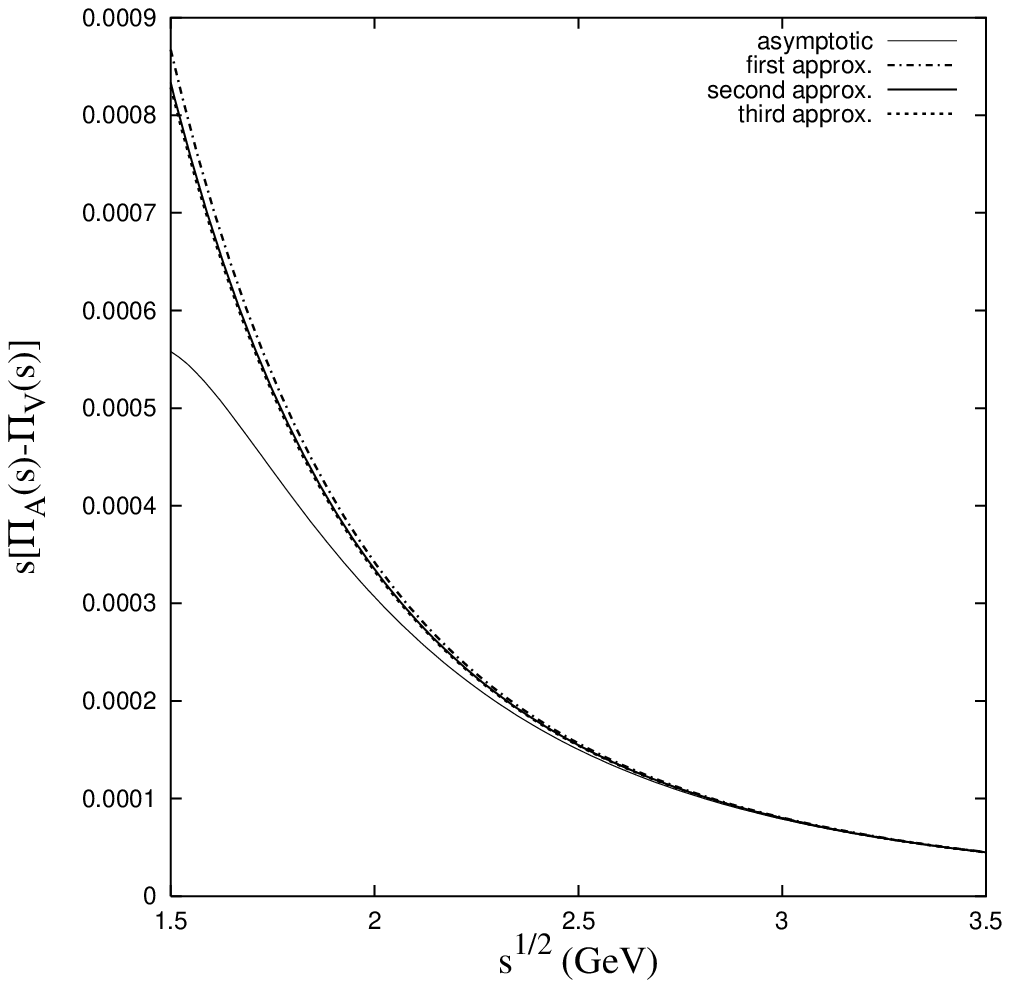}
\caption{\sl Same as Fig.4. Also shown for comparison is the two-terms
asymptotic expansion of the integrand.}
\end{center}
\end{figure}

Fig.4 shows the low energy
region $0\le\sqrt{s}\le2$ GeV. One might believe that this part will
dominate the integral, it actually turns out that the asymptotic tail makes
a non negligible contribution of approximately 20\%. It is one advantage of
this method that it introduces no error due to truncation of the integral.
One observes that approximations 2 and 3 generate curves which can hardly
be distinguished. 
Fig.5 shows a region of larger values of the integration
variable $s$ from which one can appreciate the approach to the asymptotic
regime. The two-terms asymptotic expansion is seen to
be accurate at the 10\% level for $\sqrt{s}=M_\tau$ and
becomes very accurate
provided $\sqrt{s}\gapprox2.5$ GeV. 
While the sum of
$\Pi_{A-V}^{exp}$ and $\Pi_{A-V}^{rem}$ appear to be remarkably stable they
are individually quite different from one approximation to the other. This
is illustrated in Fig.6 showing $\Pi_{A-V}^{rem}$, which is the part where the
Pad\'e interpolation procedure is used: the figure shows how this part
becomes smaller as one includes more experimental information from the
spectral functions. The curves are seen to be smooth, flat, an exhibit no
change of sign thereby justifying, a posteriori, the use of a simple rational
approximation.  

\begin{figure}[ht]
\begin{center}
\title{\bf Figure 6}
\includegraphics*[scale=1]{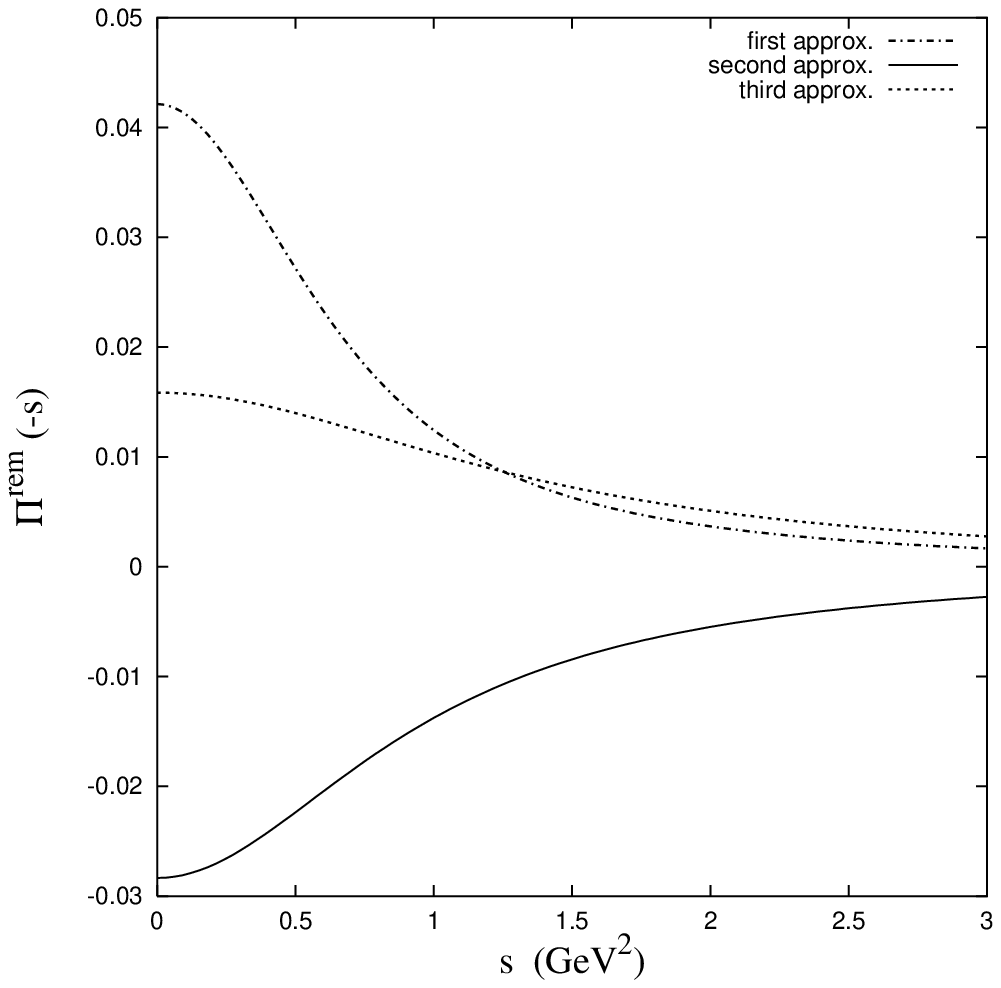}
\end{center}
\caption{\sl Plot of the remainder part $\Pi^{rem}_{A-V}(-s)$ in the sum
rule integrand (see sec.2) for the three successive approximations.}
\end{figure}

We can now perform a
stringent test of the interpolation procedure by considering the integrand
at low energy, 
comparing it with
the chiral perturbation theory expectation,
\be\lbl{limit}
\lim_{s\to0}\left[-24\pi^2
\left(\Pichir_{A-V}(-s)-{2F^2\over s}\right)+\log{s\over\mpi^2}
-{5\over3}\right]=\bar l_5\ .
\en
The low-energy constant $\bar l_5$ is known from the one-loop analysis of
the charge radius of the pion $\langle r^2\rangle^\pi_V$ 
and of the pion radiative decay
amplitude, $\pi\to e\nu\gamma$: $\bar l_5=13.1\pm 1.3$ (using $F_\pi=92.4$).
An additional $5-10\%$ uncertainty is expected from $O(\mpiq)$ contributions
to these observables, which necessitate a two-loop analysis (at present
only $\pi\to e\nu\gamma$ has been analyzed at this level of
accuracy\cite{bt}). Using our construction for $\Pi_{A-V}(-s)$ and computing
numerically the limit \rf{limit} we find,
\be\lbl{lbar5}
\bar l_5=11.8
\en
(the result differ in approx.2 and approx.3 by less than 1\%)
which is slightly smaller but compatible with the one-loop determination
quoted above. This is a rather non trivial check of the quality of the
interpolation from the low-energy domain of the chiral expansion up to the
domain of large energies, where the operator-product expansion makes sense.
The result of the sum rule evaluation of the low-energy constant $C$ are
displayed in Table 1 below: we show firstly the dimensionless quantity
$Z=C/F^4$ which is of order unity and also the quantity
$\delta\equiv(\mpidp-\mpidz-2e^2C/F^2)/(\mpidp-\mpidz)$ 
which measures the relative
importance of the subleading terms in the expansion of the $\pip-\piz$ mass
difference. 

\begin{table}[hbt]
\begin{center}
\begin{tabular}{|c|c|c|c|}\hline
\       & Approx. 1 & Approx. 2 & Approx. 3 \\ \hline
$Z$     & 0.899    & 0.854     &  0.852    \\
$\delta$&0.0177    &0.0667     &0.0683     \\ \hline
\end{tabular}
\end{center}
\caption{\sl Numerical results from the sum rule eq.\rf{zsr}
corresponding to the central values of the physical parameters. $Z$ and
$\delta$ are defined in the text.}
\label{Table 1}
\end{table}

The contribution of the subleading terms is predicted to be positive and
have a relative magnitude of $7\%$. This, of course, is 
in agreement with the upper
bound that one obtains from naive dimensional analysis of the low energy
constants, which is 20\%. What is the accuracy of this evaluation? We can
identify three sources of error: 1)the error coming from the uncertainties
in the physical parameters that enter the calculation. 2) An error coming
the chiral limit extrapolation and 3)an error associated with the assumption
that $\lambda_6$ and $\lambda_8$ are constants, which is only an
approximation.  Concerning the first source of error, we have varied all the
physical parameters independently and calculated the variation of the result
for both approximations 2 and 3. The result is shown in table 2 below.
The parameters which are not shown like $M_{\rho'}$, $\Gamma_{\rho'}$
induce very small errors. It is interesting that the individual errors are
rather different in the two approximations. For instance, the error induced
by $F$ (here the chiral limit extrapolation error was included as well) or by
$\lambda_6$, $\lambda_8$ are significantly smaller in approx. 3
than in approx. 2. This does not imply that the third approximation has a
smaller error, as it exhibits a greater sensitivity to the tail of the vector
spectral function. If one simply adds all the errors one finds very closely
the same number for the two approximations, respectively  7.3\% and 
7.5\%. 
This is suggestive that both the central value and the error are just as
reliably obtained from approx.2. In this approximation, we can also estimate
the error due
to the evaluation of the chiral limit values $\Mchir_\rho$ and
$\Gchir_\rho$. Varying the continuuum contribution in the set of sum rules
as discussed in sec. 3.2 we obtain a small contribution of 0.2\%.

\begin{table}[hbt]
\begin{center}
\begin{tabular}{|c|c|c|c|c|c|c|c|c|c|} \hline
parameters&  $F$ & $\lambda_6$ & $\lambda_8$   & $M_\rho$ & $\Gamma_\rho$&
$\beta$ & $\gamma$ & $M_{a_1}$ & $\Gamma_{a_1}$\\ \hline
error(2) &0.9 &5.1 &1.0 & 0.02 & 0.04& 0.2 &0.05 &$-$ &$-$\\
error(3) &0.04&3.4 &0.2 & 0.01 & 0.4 & 1.4 &1.5  &0.2 &0.4\\ \hline
\end{tabular}
\end{center}
\caption{\sl Percentage relative variation of the result for $C/F^2$
corresponding to the variation of the various input physical parameters
within their error bars. The second and third lines of the table correspond to
the calculation in approximation 2 and 3 respectively.}
\label{Table 2}
\end{table}

The last uncertainty arises from the assumption made so far that $\lambda_6$
and $\lambda_8$ are constants which is only true
at leading order in $\alpha_s$. This point can be investigated
quantitatively in the case of $\lambda_6$. Using the anomalous-dimension
matrix provided in ref.\cite{svz}, one can resum the leading logarithms and
obtain,
\bea\lbl{dl6}
&&\lambda_6(s)={64\pi\alpha_s(\mu)\over9} \Bigg\{
(O^a_6(\mu)+{1\over6}O^b_6(\mu))
\left[1+{9\alpha_s(\mu)\over4\pi}\log{s\over\mu^2}\right]^{-1/9}\\ \nonumber
&&+
({1\over8}O^a_6(\mu)-{1\over6}O^b_6(\mu))
\left[1+{9\alpha_s(\mu)\over4\pi}\log{s\over\mu^2}\right]^{-10/9}
+{9\alpha_s(\mu)\over32\pi}\left[{119\over6} O^a_6(\mu)+
O^b_6(\mu)\right]\Bigg\}\ .
\ena
In this expression, $O^a_6$ and $O^b_6$ are the vacuum expectation values 
of the two operators 
\bea
&&O^a_6=\langle\bar u\gamma_\mu\gamma^5{\lambda^a\over2}d
\bar d\gamma^\mu\gamma^5{\lambda^a\over2}u
-\bar u\gamma_\mu{\lambda^a\over2}d
\bar d\gamma^\mu{\lambda^a\over2}u\rangle\\ \nonumber
&&O^b_6=\langle\bar u\gamma_\mu\gamma^5 d
\bar d\gamma^\mu\gamma^5 u
-\bar u\gamma_\mu d
\bar d\gamma^\mu u\rangle
\ena
with $\lambda^a$ a color-space Gell-Mann matrix.
In principle, in order to take the s-dependence correctly into account, one
needs to know the values of both $O^a_6$ and $O^b_6$.  
However,these 
two operators are not exactly on the same footing since $O_6^b$ appears
multiplied by one
factor of $\alpha_s$ more than $O_6^a$ whenever the logarithm is not too
large. Furthermore, $O_6^b$ is suppressed in the large $N_c$ limit. A
plausible approximation, then, would be to ignore it altogether. Another
plausible approximation is that of vacuum-saturation\cite{svz},
which yields the relation,
\be\lbl{vsa}
O^b_6={3\over4} O^a_6\ .
\en
The energy dependence of $\lambda_6$ is very much suppressed in this 
approximation.
As far as the integration over the circle in the complex plane is concerned
(see eq.\rf{delta2n}), we find
that dropping the energy dependence is a very good approximation 
in any case, which does
not generate an uncertainty in the determination of $\lambda_6(M_\tau)$
larger than $1\%$. One observes from eq.\rf{dl6}
that $\lambda_6$ is a steadily decreasing function of $s$ which eventually
goes to zero when $s$ goes to infinity. Our construction can be seen as a
procedure for smoothly matching the low-to-medium and the high energy regimes
\cite{bbg}. In this sense it is clear that one must choose $\lambda_6\equiv
\lambda_6(s_0)$ where $s_0$ is the value of $s$ where the asymptotic regime
sets in, i.e. ${s_0}$ must lie between $M^2_\tau$ and $2M^2_\tau$, say, 
as can be seen from Fig.4. This
determines the constant value of $\lambda_6$ to use within $2\%$
approximately. Then, one
must take into account the contribution of the logarithms in the high energy
region of the sum rule integral. This is found to introduce a
rather small correction to the value of $C$ which ranges from $0.4$ to
$0.8\%$ depending on the hypothesis made for $O_6^b$.
In conclusion, we obtain that the overall relative error in the
determination of the parameter $C$ does not exceed 10\%.

Let us now consider the implication of this result for $O(p^4)$
low-energy parameters using the chiral expansion of the $\pip-\piz$ mass
difference at this order\cite{knecht},
\bea\lbl{dev}
&&\mpidp-\mpidz={2e^2C\over F^2}\left(1-{M^2_\pi\over16\pi^2 F^2}\left(
3\log{M^2_\pi\over\mu^2}+1\right)\right)\\ \nonumber
&&+{e^2\mpid\over16\pi^2}\left(-3\log{\mpid\over\mu^2}+4\right)
+{2\mpiq\over F^2}\left({m_d-m_u\over m_d+m_u}\right)^2 l_7+
2e^2\mpid F_k(\mu)+ O(e^4)
\ena
with
\be
F_k(\mu)= -2k^r_3(\mu)+k^r_4(\mu)+4k^r_6(\mu)+4k^r_8(\mu)\ .
\en
The $O(e^4)$ contribution  which must technically  be counted as $O(p^4)$
can be estimated to be numerically smaller 
by one order of magnitude than the $O(e^2\mpid)$
or the $O((m_u-m_d)^2)$ ones and is neglected here. From this, and
using $m_u/m_d=0.55$\cite{mu/md}, one deduces, 
\be\lbl{result}
2.2\, l_7 + F_k(M_\rho)=(-7.1\pm3.0) 10^{-2}
\en
which is our main result. For comparison,
on the basis of naive dimensional analysis
alone, one would obtain for the same quantity that it must lie in the
range $\pm 8 10^{-2}$.
The parameter $l_7$ which appears in \rf{result} is not very precisely known
but a simple resonance-saturated sum rule gives an estimate\cite{gl84}
$l_7\simeq 0.7 10^{-2}$. 

\noindent{\bf\large 5. Conclusion}

To summarize, we have attempted an evaluation of the low-energy constant
$C$ with a controlled error, on the basis of the exact sum rule expression
of Das et al.. The main practical difficulty, which is present even if
infinitely precise experimental data were available, lies in the necessity
of extrapolating the integrand to the chiral limit. A
calculational procedure was proposed in which one first 
reconstructs the relevant
current-current correlator in euclidian space making use of its smoothness
properties together with the experimental determination of two asymptotic
expansion
parameters. An approximation scheme can be developped in which one
includes spectral function components with higher and higher pion
multiplicities. This expansion was argued to converge very rapidly such
that, in practice, it is only necessary to include the one-pion and the
two-pion components. The construction of the chiral limit makes use of
recent work both on application of chiral perturbation theory to the vector
meson masses and of chiral calculations at two-loop order of current-current
correlation functions.
We have shown that under the assumption that the relative error
on the asymptotic parameter $\lambda_6$ is of the order of 10\%
(this is the actual experimental error but it does not include the
uncertainty stemming from the truncation of the OPE, which is more difficult
to evaluate), 
one can determine the parameter $C$
with an error of slightly less than 10\% and deduce a meaningful 
estimate for a combination of
subleading parameters $k_i$. These parameters are primarily useful in
calculations of radiative corrections at low energy. Another area where the
computation of the photon loop is of interest, is in relation with the
$K^+-K^0$ mass difference and the issue of Dashen's theorem violation. It is
possible that the constraint obtained here may prove useful in this context
as well.

\noindent{\bf\large Acknowledgments:} Andreas Hoecker is thanked for
communicating his thesis and for useful correspondance. This work is 
supported in part by TMR contract ERBFMRXCT980169.

\end{document}